\documentclass[aps, prl, reprint, superscriptaddress, showkeys]{revtex4-2}

\usepackage{bm}
\usepackage{graphicx}
\usepackage[title]{appendix}
\usepackage{amsmath}
\usepackage{amssymb}
\usepackage[colorlinks=true,citecolor=blue,urlcolor=blue,linkcolor=blue]{hyperref}

\usepackage{xcolor}

\begin{document}

\title{Superconducting Geometric Potential and Curvature-Enhanced Superconductivity in Curved Thin Films}

\author{Long Du}
\email{ducoollong@qq.com}
\affiliation{Guangxi University of Science and Technology, China}

\author{Qinghua Chen}
\affiliation{Guangxi University of Science and Technology, China}
% \email{chenqinghua2002@163.com}

\author{Minsi Li}
\affiliation{Guangxi University of Science and Technology, China}
% \email{liminsi@gxust.edu.cn}

\author{Jiahong Gu}
\affiliation{Guangxi University of Science and Technology, China}
% \email{jiahong\_gu@gxust.edu.cn}

\author{Guangzhen Kang}
\affiliation{Yangzhou Polytechnic Institute, China}
% \email{gzkang@nju.edu.cn}

\author{Yong-Long Wang}
\email{wangyonglong@lyu.edu.cn}
\affiliation{School of Physics and Electronic Engineering, Linyi University, China}

\begin{abstract}
The impact of pure geometric curvature on the superconductivity of thin films remains controversial due to the masking effects of mechanical strain.
To isolate the purely geometric effects, we derive the linearized Ginzburg-Landau (GL) equation for a curved ultra-thin superconducting film in the presence of a magnetic field.
By introducing a novel transverse order parameter that varies slowly along the film with the superconducting/vacuum boundary condition, we decouple the linearized GL equation into a transverse component and a surface component in the thin-layer quantization scheme.
A superconducting geometric potential (GP) is present in the surface equation, which can substantially affect the nucleation of the superconducting state in the curved ultra-thin superconducting film.
From the perspective of the GL free energy, the superconducting GP reduces the coefficient of the quadratic term of the order parameter, which enables the curved film to stay in the superconducting state even when the superconducting parameter $\alpha$ becomes positive.
Based on our equivalent equation, for a superconducting thin film with uniform curvature, the relative increase of the critical temperature is proportional to the magnitude of the superconducting GP.
As an example, we numerically investigate the phase transition of a rectangular superconducting film bent around a cylindrical surface, and the numerical results are in good agreement with the theoretical expectations.
We further propose a strain-free experimental validation using ultracold atomic condensates, where nested superfluid shells enforce Neumann boundary conditions to isolate the superconducting GP.
\end{abstract}


\maketitle

\section{Introduction \label{sec:Introduction}}

The ability to structure superconductors at the mesoscopic scale has opened up a rich landscape of quantum phenomena determined by their geometry and topology.
Rapid advances in micro- and nanofabrication, such as the recent focused ion beam induced deposition (FIBID) of complex three-dimensional (3D) heterostructures \cite{Allen2025NewDirections,lutchyn2018majorana}, have made it possible to engineer boundary conditions that fundamentally alter the superconducting condensate \cite{moshchalkov1994quantum,moshchalkov2000handbook,guo2004superconductivity}.
When sample dimensions approach the coherence length, spatial confinement leads to exotic behaviors distinct from the bulk: from the seminal observation of quantum periodicity in transition temperatures \cite{Little1962Observation} and the fine-tuning of frustration in 2D networks \cite{pannetier1984experimental}, to topology-dependent phase boundaries \cite{moshchalkov1995effect} and the emergence of bistable states \cite{Geim1997Phase}, alongside the nucleation of surface superconductivity or giant vortices \cite{saint1965etude,fink1966magnetic,cren2011vortex}.
Moreover, the interplay between confinement and discrete symmetry enforces the spontaneous formation of vortex-antivortex molecules in regular polygons \cite{chibotaru2000symmetry,Misko2003Stable}, such as squares \cite{Bonca2001Phase,Melnikov2002Vortex} and triangles \cite{Chibotaru2001Vortex}, which represents a symmetry-restoring phenomenon strictly forbidden in macroscopic samples \cite{Chen2009Vortex}.

The geometric curvature of quasi-2D systems constitutes another fundamental degree of freedom beyond confinement.
The effective quantum mechanics of particles confined to curved surfaces is conventionally obtained by using the thin-layer quantization scheme (TLQS) \cite{Jensen1971Quantum,DaCosta1981Quantum},
where the physical limit of an infinite confining potential yields an attractive geometric potential (GP) dependent on the mean and Gaussian curvatures \cite{Burgess1993Fermions,Jensen2009Quantum}.
Although this curvature-induced potential \cite{Schuster2003Quantum,Ferrari2008Schrdinger} and the associated geometric momentum \cite{Liu2013Geometric,Wang2014Pauli} have been verified in diverse systems, ranging from the Riemannian geometric effects in peanut-shaped $\mathrm{C}_{60}$ polymers \cite{Onoe2012Observation} to transport in photonic topological crystals \cite{Szameit2010Geometric} and geometric momenta in plasmonic waveguides \cite{aoki2001electronic,Spittel2015Curvature}, its role in superconductivity remains controversial.
Recent research has definitively demonstrated that geometric effects drive dynamics in chiral superfluids, inducing a ``geometric Meissner effect'' and a ``curvature anomaly'' \cite{bai2025geometry}.
However, in solid-state superconductors, distinguishing pure curvature effects from those of mechanical strain poses a significant challenge \cite{bozovic2002epitaxial,ahadi2019enhancing}.
Strain is known to renormalize the GP by orders of magnitude \cite{Ortix2011Curvature,ortix2010effect} or induce Rashba spin-orbit interactions \cite{gentile2013curvature}, typically masking the intrinsic geometric effects.
Strikingly, recent theoretical work suggests that for conventional $s$-wave superconductors, the impact of pure geometric curvature on the order parameter is negligible once strain is eliminated \cite{Heinrich2025Bending}.
This raises a fundamental question: Can pure geometric curvature, independent of strain and exotic chirality, provide a sufficiently strong mechanism to significantly enhance conventional superconductivity?

To resolve this controversy, we revisit the problem within the Ginzburg-Landau (GL) theory \cite{Tinkham1996}.
This framework has proven exceptionally robust in capturing topology-driven phenomena across diverse manifolds: from persistent currents and Little-Parks oscillations in rings and hollow cylinders \cite{carillo2010little,schwiete2009persistent,gladilin2012aharonov,carapella2011single}, to topological nodal states in M\"{o}bius strips \cite{hayashi2001little,hayashi2005superconductivity}, and complex vortex dynamics in helices \cite{fomin2017superconducting,cordoba2019three}, curved strips \cite{sabatino2011magneto,rezaev2016branching}, and spherical shells \cite{gladilin2008vortices,gladilin2012negative}.
Within the TLQS framework, we derive an effective linearized GL equation for curved thin films with the superconducting/vacuum (S/V) boundary conditions \cite{Baelus2006One,moshchalkov2009type}.
Our central finding is the emergence of a novel superconducting GP, $V_\mathrm{g}^\mathrm{s} = -\hbar^2 / (2\mu) (2M^2 - K)$, with $M$ and $K$ being the mean and Gaussian curvatures of superconducting films, respectively.
In contrast to the GP \cite{DaCosta1981Quantum}, the superconducting GP arises from the coupling between the surface curvature and the transverse Neumann boundary condition.
We also demonstrate that $V_\mathrm{g}^\mathrm{s}$ serves as an effective attractive interaction lowering the GL free energy, which favors the nucleation of the superconducting state.
Consequently, unlike perspectives focusing on strain-induced effects \cite{Heinrich2025Bending}, the present GL framework analysis reveals that pure geometry can indeed provide a robust mechanism for the substantial enhancement of the critical temperature $T_\mathrm{c}$.

This paper is organized as follows.
In Sec.\ \ref{sec:Linearized_GL}, we derive the effective 2D linearized GL equation and obtain the superconducting GP by decoupling the transverse and surface components of the order parameter via the gauge transformation \cite{Ferrari2008Schrdinger, FleckingerPell1995GaugesFT}.
In Sec.\ \ref{sec:GL_free}, we analyze the mechanism of superconductivity enhancement from the perspective of GL free energy and establish a direct relation between the curvature and the increase in $T_\mathrm{c}$.
In Sec.\ \ref{sec:Curvature-induced_nucleation}, we numerically validate the effective linearized GL equation by simulating a rectangular film bent around a cylinder in a perpendicular magnetic field \cite{tanaka1993epitaxial,Schulz1994}, demonstrating that the $T_\mathrm{c}$ enhancement scales quadratically with the mean curvature.
In Sec.\ \ref{sec:Proposed_Experimental}, we propose a strain-free validation platform using ultracold atomic condensates, demonstrating how a nested-shell geometry enforces the requisite Neumann boundary conditions to provide a smoking-gun signature of the superconducting GP.
Finally, conclusions are given in Sec.\ \ref{sec:Conclusion}.

\section{Linearized GL equation on a curved thin superconducting film \label{sec:Linearized_GL}}
We first consider the linearized GL equation, which accurately describes superconducting nucleation near the phase transition.
Since the response to an external magnetic field is fundamental to characterizing mesoscopic superconductors and determining their critical parameters \cite{Tinkham1996, moshchalkov2009type}, incorporating the vector potential is essential for analyzing superconductivity in curved geometries.
In the presence of a magnetic field $\bm{B}=\nabla\times\bm{A}$, where $\bm{A}$ is the vector potential, the stationary linearized GL equation is given by \cite{Tinkham1996, Baelus2006One}:
\begin{equation}\label{eq:LGL}
  \frac{1}{2\mu }\left(-\mathrm{i}\hbar\nabla-\frac{Q}{c}\bm{A}\right)^2\psi=-\alpha\psi,
\end{equation}
where $\mu$ and $Q$ denote the effective mass and charge of a Cooper pair, respectively.
Here, $\psi$ is the superconducting order parameter, with $|\psi|^2$ proportional to the Cooper-pair density.
The GL parameter $\alpha$ is defined as:
\begin{equation}
  -\alpha = \frac{\hbar^2}{2\mu \xi_0^2} \left( 1 - \frac{T}{T_\mathrm{c}} \right),
\end{equation}
where $\xi_0$ is the zero-temperature coherence length, $T$ is the temperature, and $T_\mathrm{c}$ represents the bulk critical temperature in the absence of a magnetic field.
Equation (\ref{eq:LGL}) is invariant under the gauge transformation \cite{FleckingerPell1995GaugesFT}:
\begin{equation}\label{eq:gauge_invariance}
  \left(\psi,\bm{A}\right) \mapsto \left(\psi^\prime,\bm{A}^\prime\right)
  = \left(\psi \mathrm{e}^{\mathrm{i} Q \gamma/\hbar}, \bm{A}+\nabla\gamma \right),
\end{equation}
where $\gamma$ is an arbitrary scalar function.

\begin{figure}[!h]\centering
  \includegraphics[width=0.6\columnwidth]{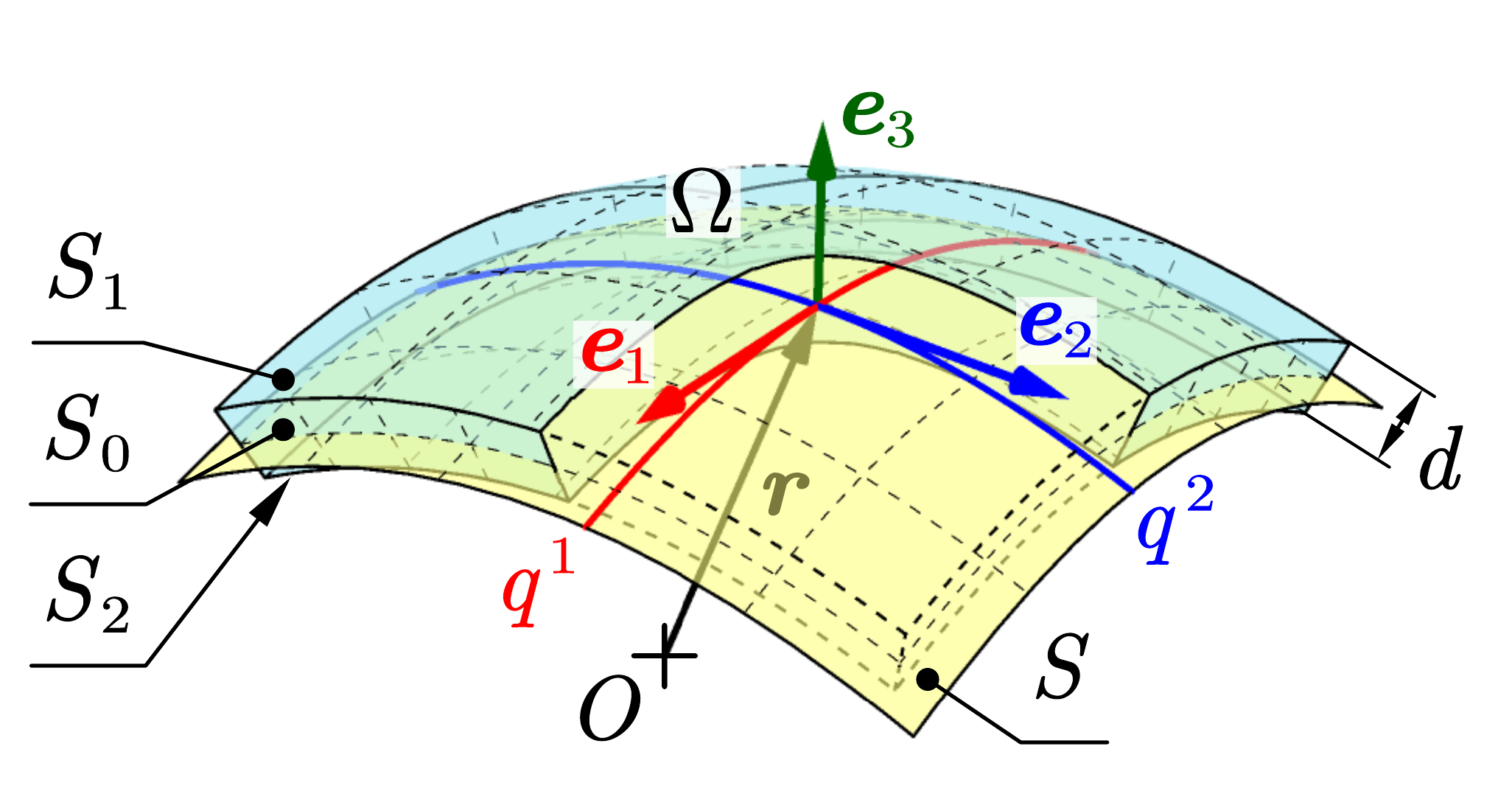}
  \caption{\label{fig:curved_surface_general}
    (Color online) Schematic representation of a curved superconducting film $\Omega$ with uniform thickness $d$, bounded by surfaces $S_1, S_2$ and sidewall $S_0$.
    A point on the central surface $S$ is parameterized by $\bm{r}(q^1, q^2)$, where $q^1, q^2$ are curvilinear coordinates.
    Vectors $\bm{e}_1$ and $\bm{e}_2$ form the tangent basis on $S$, while $\bm{e}_3$ is the unit normal vector.
  }
\end{figure}

A curved superconducting film of uniform thickness $d$ is considered, as illustrated in Fig.\ \ref{fig:curved_surface_general}, that is embedded in an insulating medium or vacuum.
The superconducting region, denoted by $\Omega$, is bounded by two parallel surfaces, $S_1$ and $S_2$, representing the top and bottom interfaces, and the side surface $S_0$.
We define a central surface $S$ equidistant from $S_1$ and $S_2$, and assume $S_0$ is locally perpendicular to $S$ at their intersection.
The physical boundary condition requires the normal component of the supercurrent to vanish at the insulating interfaces:
\begin{equation} \label{eq:BC}
  \left[ \bm{n} \cdot \left( \mathrm{i}\hbar \nabla + Q\bm{A} / c \right) \psi \right] \big|_{\partial \Omega}=0,
\end{equation}
where $\bm{n}$ is the outward unit normal vector to the total boundary $\partial \Omega = S_0 \cup S_1 \cup S_2$.

In the thin-film limit, where the thickness $d$ is much smaller than the minimum principal radius of curvature of the surface, Cooper pairs within $\Omega$ are effectively confined to the vicinity of the surface $S$.
Following the TLQS developed by da Costa \cite{DaCosta1981Quantum}, we parameterize the position vector $\bm{R}$ in $\Omega$ as:
\begin{equation}\label{eq:position_Omega_S}
  \bm{R} ( q^1, q^2, q^3 ) = \bm{r} ( q^1, q^2 ) + q^3 \bm{e}_3 ( q^1, q^2 ),
\end{equation}
where $\bm{r}(q^1, q^2)$ locates a point on $S$, $\bm{e}_3$ is the unit normal vector to $S$, and $q^3 \in [-d/2, d/2]$ represents the coordinate along the normal direction.
Indices $i,j,k=1,2,3$ denote spatial components in $\Omega$, while $a,b=1,2$ refer to surface components on $S$; Einstein summation convention is adopted.
The basis vectors in $\Omega$ are given by $\partial_i \bm{R}$. Specifically, the tangent basis vectors read
\begin{equation}
  \bm{\epsilon}_a = \partial_a \bm{R} = \partial_a \bm{r} + q^3 \partial_a \bm{e}_3,
\end{equation}
while the normal basis vector is $\bm{e}_3 = \bm{e}_1 \times \bm{e}_2 / |\bm{e}_1 \times \bm{e}_2|$, where $\bm{e}_a = \partial_a \bm{r}$.
The derivatives of the normal vector satisfy the Weingarten equation:
\begin{equation}\label{eq_wgt}
  \partial_a \bm{e}_3 = w_a^b \bm{e}_b,
\end{equation}
where $w_a^b$ are the components of the Weingarten curvature matrix.
The covariant metric tensor components in $\Omega$ ($G_{ij}$) and on $S$ ($g_{ab}$) are related via Eq.\ (\ref{eq:position_Omega_S}) as:
\begin{gather}
  G_{ab} = g_{ab} + \left( \hat{w} \hat{g} + \hat{g}^T \hat{w}^T \right)_{ab} q^3+ \left( \hat{w} \hat{g} \hat{w}^T \right)_{ab} (q^3)^2, \nonumber\\
  G_{a3}=G_{3a}=0,\ G_{33}=1,
  \label{eq:relation_metric}
\end{gather}
where $\hat{w}$ and $\hat{g}$ denote the matrices with components $w_a^b$ and $g_{ab}$, respectively.
The metric determinants $G = \det(G_{ij})$ and $g = \det(g_{ab})$ satisfy \cite{DaCosta1981Quantum}:
\begin{equation}\label{eq:G}
  G=f^2 g,
\end{equation}
where $f = 1 + 2M q^3 + K (q^3)^2$, with $M = \mathrm{Tr}(\hat{w})/2$ and $K = \det(\hat{w})$ representing the mean and Gaussian curvatures of $S$, respectively \cite{Ferrari2008Schrdinger, docarmo2016}.

The linearized GL equation can be expressed in general curvilinear coordinates using the covariant Laplacian.
Substituting the metric relations into Eq.\ (\ref{eq:LGL}), we obtain:
\begin{align}
  & \frac{1}{2\mu} \Bigg[
    -\frac{\hbar^2}{\sqrt{G}} \partial_i \left(\sqrt{G} G^{ij} \partial_j \psi \right)
    + \frac{2\mathrm{i} \hbar Q}{c} G^{ij} A_i \partial_j \psi
  \nonumber\\
  & + \frac{\mathrm{i} \hbar Q}{c \sqrt{G}} \partial_i \left( \sqrt{G} G^{ij} A_j \right) \psi
    + \frac{Q^2}{c^2} G^{ij} A_i A_j \psi
  \Bigg] = -\alpha \psi.
  \label{eq:LGL_curv}
\end{align}
Similarly, the boundary condition (\ref{eq:BC}) becomes:
\begin{equation}\label{eq:BC_curv}
  \left. \left\{ \bm{n}\cdot\left[ G^{ij}\left( \mathrm{i}\hbar \,\partial_i + Q A_i / c \right) \psi \,\partial_j \bm{R} \right] \right\} \right|_{\partial\Omega} = 0.
\end{equation}
To decouple the surface and transverse order parameter in the thin-film limit, we introduce a rescaled order parameter $\chi$ defined by $\psi = f^{-1/2}\chi$.
This transformation, standard in the TLQS, ensures the conservation of probability density on the surface \cite{DaCosta1981Quantum, Liu2013Geometric}.
Taking the limit $d \to 0$ (i.e., $q^3 \to 0$), the derivatives of $\psi$ relate to $\chi$ as:
\begin{align}
  & \lim_{q^3\to 0} \psi = \chi, \ \lim_{q^3\to 0} \partial_3 \psi = \partial_3 \chi - M \chi, \nonumber\\
  & \lim_{q^3\to 0} \partial_3^2 \psi = \partial_3^2 \chi - 2M \partial_3 \chi + (3 M^2 - K) \chi.
  % \label{eq:Limit}
\end{align}
Substituting these limits into Eq.\ (\ref{eq:LGL_curv}), the linearized GL equation can be expanded as:
\begin{align}
  & \frac{\hbar^2}{2\mu} \left[
    -\frac{1}{\sqrt{g}} \partial_a \left( \sqrt{g} g^{ab} \partial_b \chi \right)
    - \partial_3^2 \chi \right]
  + \frac{\mathrm{i} \hbar Q}{2 \mu c} \bigg[
    \frac{1}{\sqrt{g}} \partial_a ( \sqrt{g} \nonumber\\
  &   \times g^{ab} A_b ) \chi
    + \partial_3 A_3 \chi
    + 2 (g^{ab} A_a \partial_b \chi + A_3 \partial_3 \chi) \bigg] \nonumber\\
  & + \frac{Q^2}{2\mu c^2} \left( g^{ab} A_a A_b + A_3^2 \right) \chi
  + V_\mathrm{g} \chi = - \alpha \chi,
  \label{eq:GL_curv_1}
\end{align}
where $V_\mathrm{g} = -\hbar^2 (M^2 - K) / (2\mu)$ is the GP for a spinless particle \cite{DaCosta1981Quantum}.
Correspondingly, the boundary conditions (\ref{eq:BC_curv}) split into:
\begin{align}
  & \left\{ \bm{n} \cdot \left[ g^{ab} \left( \mathrm{i}\hbar \partial_a\chi + (Q/c) A_a\chi \right) \bm{e}_b \right] \right\} \big|_{S_0} =0,   \label{eq:BC_curv_surface} \\
  & \left[ \mathrm{i}\hbar \left(\partial_3 \chi - M \chi \right) + (Q/c) A_3 \chi \right] \big|_{S_1 \cup S_2}=0.
  \label{eq:BC_curv_transverse}
\end{align}

The term $A_3 \partial_3 \chi$ in Eq.\ (\ref{eq:GL_curv_1}) couples the transverse and surface components of the GL equation.
To eliminate this coupling, we adopt the gauge transformation method introduced by Ferrari and Cuoghi \cite{Ferrari2008Schrdinger}, utilizing the scalar function (Eq.\ (\ref{eq:gauge_invariance})):
\begin{equation}\label{eq:gauge_suit}
  \gamma(q^1,q^2,q^3) = - \int_{0}^{q^3} A_3 ( q^1, q^2, z ) \,\mathrm{d}z.
\end{equation}
In the limit $ q^3 \to 0 $, we have $ A^\prime_3=0 $ and $ \partial_3 A_3^\prime=0 $, while $ A_1 $ and $ A_2 $ remain unchanged.
In the TLQS, which restricts transverse fluctuations to depend solely on $q^3$ \cite{DaCosta1981Quantum}, a transverse Neumann-type constraint typically renders Schr\"{o}dinger-type equations on surfaces nonseparable if the mean curvature is non-uniform \cite{Jensen2009Quantum}.
To derive a surface linearized GL equation that accommodates the transverse boundary conditions in Eq.\ (\ref{eq:BC_curv_transverse}), we introduce a decomposed order parameter:
\begin{equation}\label{eq:hypothesis}
  \chi(q^1, q^2, q^3) = \chi_\mathrm{s}(q^1,q^2) \,\chi_\mathrm{t}(q^1,q^2,q^3),
\end{equation}
where $\chi_\mathrm{s}$ and $\chi_\mathrm{t}$ represent the surface and transverse components of the order parameter, respectively.
It is straightforward to verify that the separability condition in TLQS holds when $\chi_\mathrm{t}$ is a slowly varying function along $S$, ensuring $\lim_{q^3\to 0}\partial_a \chi_\mathrm{t} = 0$ \cite{Jensen2009Quantum}.
While the strict variable separation $\chi = \chi_s \chi_t$ holds exactly for surfaces with uniform mean curvature such as cylinders and spheres, it also serves as a robust adiabatic approximation for surfaces with spatially varying curvature. Specifically, provided the curvature varies smoothly under the condition $|\nabla_s M| \ll M/\xi_0$, the cross-derivative terms can be safely neglected.

Applying the gauge transformation (\ref{eq:gauge_suit}) and decomposing the eigenvalue as $\alpha = \alpha_\mathrm{s} + \alpha_\mathrm{t}$ with $\alpha_\mathrm{s}$ and $\alpha_\mathrm{t}$ corresponding to the surface and transverse eigenstates, respectively, we decouple Eq.\ (\ref{eq:GL_curv_1}) into two independent equations:
\begin{equation}
  -\frac{\hbar^2}{2\mu} \partial_3^2 \chi_\mathrm{t} = -\alpha_\mathrm{t} \chi_\mathrm{t},
  \label{eq:LGL_curv_transverse}
\end{equation}
and
\begin{equation}
  \frac{1}{2\mu} \left(-\mathrm{i}\hbar \nabla_\mathrm{s} - \frac{Q}{c} \bm{A}_\mathrm{s}\right)^2 \chi_\mathrm{s} + V_\mathrm{g} \chi_\mathrm{s} = -\alpha_\mathrm{s} \chi_\mathrm{s},
  \label{eq:LGL_curv_surface}
\end{equation}
where $ \nabla_\mathrm{s} = g^{ab} \bm{e}_a \partial_b $ and $ \bm{A}_\mathrm{s} = g^{ab}\bm{e}_a A_b $.
Eq.\ (\ref{eq:LGL_curv_transverse}) governs the transverse confinement along the normal direction, while Eq.\ (\ref{eq:LGL_curv_surface}) describes the effective superconducting state on the curved surface $S$.
The transverse component of the boundary condition derived in Eq.\ (\ref{eq:BC_curv_transverse}) takes the form:
\begin{equation}\label{eq:bc4_normal}
  \left[ (\partial_3 - M) \chi_\mathrm{t} \right] \big|_{q^3 = \pm d/2} = 0.
\end{equation}
From Eq.\ (\ref{eq:LGL_curv_transverse}), the permissible eigenstate satisfying the boundary condition (\ref{eq:bc4_normal}) is found to be $\chi_\mathrm{t} \sim \exp(M q^3)$, yielding the corresponding eigenvalue $-\alpha_\mathrm{t} = -\hbar^2 M^2 / (2\mu)$ (see Appendix \ref{app:Exact_Solution}).
Consequently, under the surface boundary condition (\ref{eq:BC_curv_surface}), the total eigenvalue $-\alpha$ is determined by computing the lowest eigenvalue of the following effective surface equation:
\begin{equation}\label{eq:LGL_curv_surface1}
  \frac{1}{2\mu} \left(-\mathrm{i}\hbar \nabla_\mathrm{s} - \frac{Q}{c} \bm{A}_\mathrm{s}\right)^2 \chi_\mathrm{s} + V_\mathrm{g}^\mathrm{s} \chi_\mathrm{s} = -\alpha \chi_\mathrm{s},
\end{equation}
where we identify the superconducting GP as:
\begin{equation} \label{eq:supGP}
  V_\mathrm{g}^\mathrm{s} = V_\mathrm{g} - \alpha_\mathrm{t} = -\frac{\hbar^2}{2\mu} (2M^2 - K).
\end{equation}
This potential is specific to thin superconducting films and is strictly negative on curved surfaces.

As a fundamental conclusion of this derivation: \emph{provided the newly introduced transverse order parameter varies sufficiently slowly along the surface, the linearized GL equation can be analytically decoupled into transverse and surface components under the S/V boundary condition (\ref{eq:BC_curv_surface}), with the superconducting nucleation governed by Eq.\ (\ref{eq:LGL_curv_surface1}).}
The film thickness should generally be much smaller than $\xi_0$ to ensure that the transverse component of the order parameter remains in the ground state, thereby validating the use of Eq.\ (\ref{eq:LGL_curv_surface1}).
Furthermore, the radius of curvature of the film should be significantly larger than $\xi_0$.
This condition not only minimizes strain within the film, allowing geometric curvature to dominate the order parameter's behavior, but also mitigates the risk of film fracture.

\section{GL free energy and the mechanism of curvature-enhanced superconductivity \label{sec:GL_free}}
In this section, we analyze the impact of the superconducting GP on the nucleation of the superconducting state from the perspective of the GL free energy.
Within the framework of the GL theory, the total free energy difference between the superconducting and normal states for the film illustrated in Fig.\ \ref{fig:curved_surface_general} is given by:
\begin{equation}\label{eq:free_energy}
  F = F_\mathrm{n} + \int_\Omega \left( \psi^* \hat{L} \psi + \alpha |\psi|^2 + \frac{\beta}{2} |\psi|^4 + \varepsilon_\mathrm{mag} \right) \mathrm{d}V,
\end{equation}
where $F_\mathrm{n}$ represents the free energy of the normal state, $\beta$ is the phenomenological GL parameter, and $\varepsilon_\mathrm{mag}$ denotes the magnetic field energy density.
Here, $\hat{L} = (2\mu)^{-1} [-\mathrm{i}\hbar \nabla - (Q / c) \bm{A}]^2$ is the kinetic energy operator \cite{Chen2009Vortex}.
Utilizing the curvilinear coordinates defined in Sec.\ \ref{sec:Linearized_GL} and substituting the decoupled order parameter ansatz (Eq.\ (\ref{eq:hypothesis})) along with the gauge transformation (Eq.\ (\ref{eq:gauge_suit})), we can effectively reduce Eq.\ (\ref{eq:free_energy}) to a surface integral in the thin-film limit:
\begin{align}
  & F = F_\mathrm{n} + \Lambda \int_S \bigg( \chi_\mathrm{s}^* \hat{L}_\mathrm{s} \chi_\mathrm{s} + (\alpha + V_\mathrm{g}^\mathrm{s}) |\chi_\mathrm{s}|^2 \nonumber \\
  & \quad + \frac{\beta}{2} |\chi_\mathrm{s}|^4 + \varepsilon_\mathrm{mag}^\mathrm{s} \bigg) \,\mathrm{d}{s},
  \label{eq:free_energy_2d}
\end{align}
where the coefficient $\Lambda$ arises from the integration over the film thickness, $\hat{L}_\mathrm{s} = (2\mu)^{-1} [ -\mathrm{i}\hbar \nabla_\mathrm{s} - (Q / c) \bm{A}_\mathrm{s} ]^2$ is the surface kinetic energy operator, $\mathrm{d}s = \sqrt{g} \,\mathrm{d}q^1 \,\mathrm{d}q^2$ is the area element, and $\varepsilon_\mathrm{mag}^\mathrm{s}$ represents the surface magnetic energy density.
Equation (\ref{eq:free_energy_2d}) explicitly shows that the superconducting GP $V_\mathrm{g}^\mathrm{s}$, which encapsulates the effects of both curvature and the transverse Neumann boundary condition, acts as an effective attractive potential.
Since $V_\mathrm{g}^\mathrm{s}$ is negative, it reduces the coefficient of the quadratic term $|\chi_\mathrm{s}|^2$ in the free energy expansion.
Consequently, the condition for the superconducting phase transition is relaxed, allowing the film to maintain the superconducting state even when the bulk parameter $\alpha$ becomes positive (i.e., at temperatures slightly above the bulk $T_\mathrm{c}$).
For a flat superconducting film where $V_\mathrm{g}^\mathrm{s}$ vanishes, the critical temperature $T^\ast$ is determined by the standard relation:
\begin{equation}
  -\alpha_0 = \frac{\hbar^2}{2 \mu \xi_0^2} \left( 1 - \frac{T^\ast}{T_\mathrm{c}} \right),
\end{equation}
where $-\alpha_0$ corresponds to the lowest eigenvalue of the surface kinetic operator in Eq.\ (\ref{eq:LGL_curv_surface1}) without the GP.
When the film is bent into a configuration with uniform curvature, the new critical temperature $\tilde{T}^\ast$ is determined by the lowest eigenvalue of the modified operator $-(\alpha_0 + V_\mathrm{g}^\mathrm{s})$.
This leads to the following relation between $\tilde{T}^\ast$ and $T^\ast$:
\begin{equation}
  \label{eq:Tc_mod}
  \frac{\tilde{T}^\ast - T^\ast}{T_\mathrm{c}} = \xi_0^2 \left( 2M^2 - K \right).
\end{equation}
This result leads to the fundamental conclusion: \emph{in comparison with a flat film, the critical temperature of a curved superconducting film is enhanced by a factor proportional to $\xi_0^2 (2M^2 - K)$.}
This curvature-induced enhancement can be qualitatively understood within the microscopic BCS framework by considering the center-of-mass motion of Cooper pairs.
In the derivation of the GL theory from the microscopic BCS theory \cite{gor1959microscopic}, the order parameter $\psi$ represents the center-of-mass wavefunction of Cooper pairs.
The kinetic energy term in the GL equation accounts for the energy cost associated with the spatial confinement and variation of $\psi$.
In a curved thin film, the strong quantum confinement in the normal direction imposes a constraint on the transverse component of the pair wavefunction.
Unlike in a flat geometry, the interplay between the curved metric and the Neumann boundary condition effectively modifies the transverse kinetic energy contribution.
Specifically, the negative sign of $V_\mathrm{g}^\mathrm{s}$ implies a reduction in the transverse ground-state kinetic energy required to confine Cooper pairs within the curved film.
This reduction in kinetic energy cost acts equivalently to an attractive potential, stabilizing the superconducting condensate at higher temperatures, distinct from any strain-induced modification of the electron-phonon interaction strength.

\section{Curvature-induced nucleation in a cylindrical superconducting film \label{sec:Curvature-induced_nucleation}}

Although the continuous formulation of the GL theory is intrinsically limited by strong thermal fluctuations (e.g., BKT transitions) and quantum size effects in the extreme thin-film limit ($d \to 0$), recent experiments on 2D van der Waals superconductors, such as few-layer NbSe$_2$ \cite{xi2016ising}, have demonstrated that robust macroscopic superconducting coherence persists down to nanometer thicknesses. Consequently, our phenomenological GL framework remains highly applicable as an effective macroscopic model, provided the system is safely outside the critical fluctuation regime of the BKT transition.

\begin{figure}[!h]\centering
  \includegraphics[width=0.6\columnwidth]{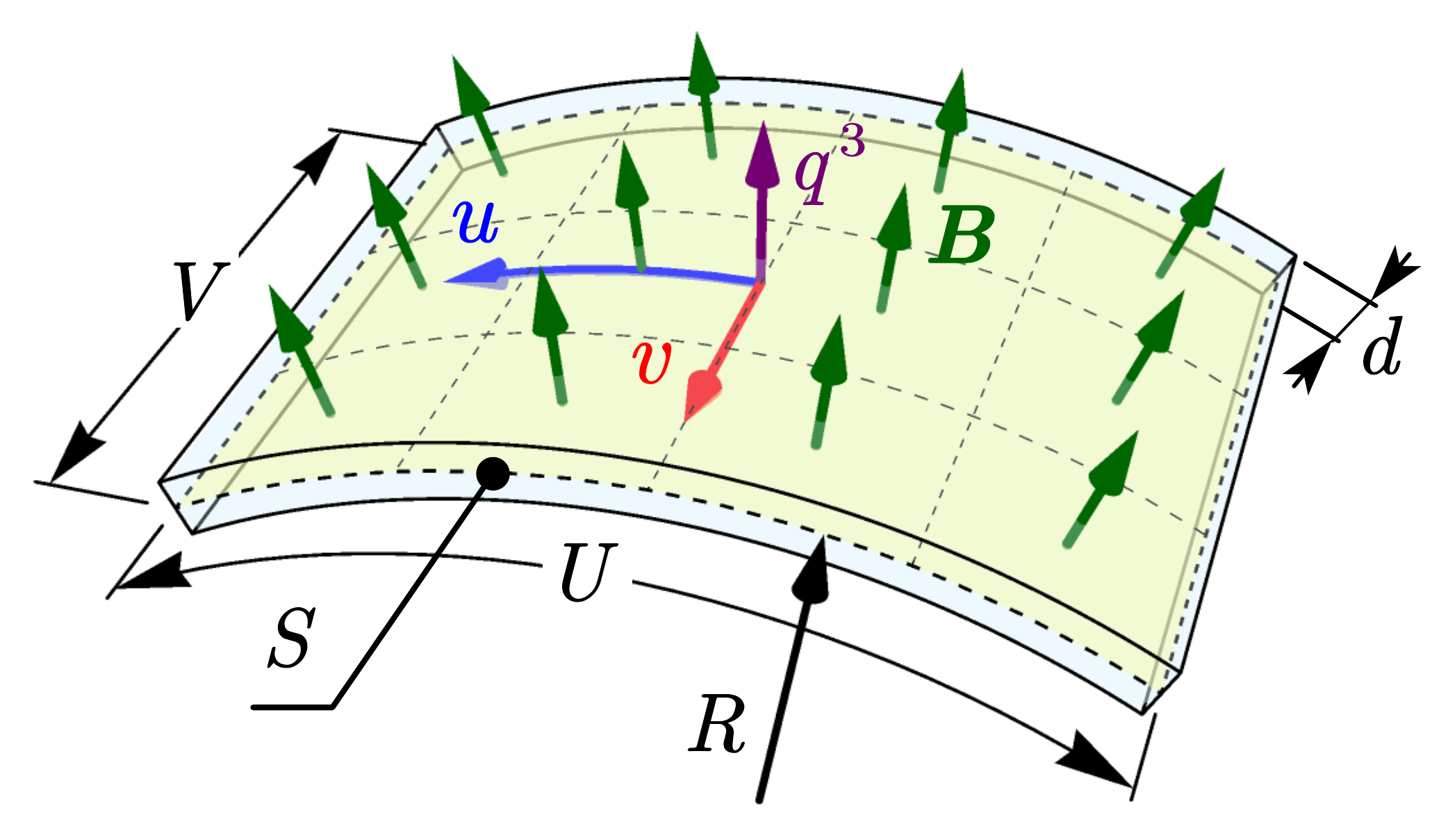}
  \caption{\label{fig:cylindrical_segment}(Color online)
    A rectangular thin superconducting film of width $U$, height $V$, and thickness $d$, bent along its width to conform to a cylindrical surface of radius $R$.
    The stress-free central surface $S$ is parameterized by orthogonal coordinates $(u,v)$ aligned with the azimuthal and axial directions, respectively.
    Coordinate $q^3$ denotes the normal distance from $S$.
    An external magnetic field $\bm{B}$ (dark green arrows) is applied perpendicular to the film surface with uniform intensity.
  }
\end{figure}

To quantitatively investigate the geometric effects on the normal-superconducting phase transition, we numerically investigate the nucleation process of a rectangular superconducting thin film bent around a cylindrical surface.
For computational convenience, we introduce the following dimensionless variables:
\begin{gather}
  \xi = \xi_0 \xi^\prime,\
  q^a = \xi_0 {q'}^a,\
  \partial_a = \xi_0^{-1} \partial^\prime_a,\
  M = M'/\xi_0, \nonumber\\
  K = K'/\xi_0^2,\
  \bm{A} = \frac{\hbar c}{\xi_0 Q}\bm{A}',\
  \alpha = \frac{\hbar^2}{2\mu \xi_0^2}\alpha',
  \label{eq:dim_trans}
\end{gather}
where the primed quantities denote the dimensionless variables.
By applying the algebraic transformations in Eq.\ (\ref{eq:dim_trans}) and dropping the primes for simplicity, we obtain the scaled versions of Eqs.\ (\ref{eq:LGL_curv_surface1}) and (\ref{eq:BC_curv_surface}) as follows:
\begin{align}
   & -\frac{1}{\sqrt{g}} \partial_a \left( \sqrt{g} g^{ab} \partial_b \chi_\mathrm{s} \right)
  + \frac{\mathrm{i}}{\sqrt{g}} \partial_a \left( \sqrt{g} g^{ab} A_b \right) \chi_\mathrm{s} \nonumber \\
   & + 2 \mathrm{i} g^{ab} A_a \partial_b \chi_\mathrm{s}
  + g^{ab} A_a A_b \chi_\mathrm{s}
  + V_\mathrm{g}^\mathrm{s} \chi_\mathrm{s}
  = -\alpha \chi_\mathrm{s},
  \label{eq:LGL_curv_surface_scaled}
\end{align}
and
\begin{equation}
  \left\{ \bm{n} \cdot \left[ g^{ab} (\mathrm{i} \partial_a + A_a ) \chi_\mathrm{s} \, \bm{e}_b \right] \right\} \big|_{S_0}=0,
  \label{eq:bc_curv_surf_scaled}
\end{equation}
where the dimensionless superconducting GP is $V_\mathrm{g}^\mathrm{s} = -(2M^2 - K)$, and the scaled GL parameter is given by $-\alpha = 1 - T/T_\mathrm{c}$.
Specifically, we consider a rectangular superconducting film with width $U$, height $V$, and thickness $d$, bent along its width to conform to the surface of a cylinder, as illustrated in Fig.\ \ref{fig:cylindrical_segment}.
The stress-free central surface $S$, with radius $R$, is parameterized by two dimensionless orthogonal coordinates $(u,v)$, where $u$ and $v$ correspond to the circumferential (width) and axial (height) directions, respectively.
To investigate the intrinsic impact of curvature on superconductivity, we apply a magnetic field $\bm{B}$ that is locally perpendicular to the film surface with uniform intensity.
Although a uniform external magnetic field is experimentally easy to implement, it is unsuitable here as it results in a reduced average normal field, which would trivially enhance $T_\mathrm{c}$ and obscure the effect of the superconducting GP.
Experimentally, such a perpendicular field distribution can be realized by attaching the superconducting film to a magnetic substrate with perpendicular magnetization \cite{tanaka1993epitaxial,Schulz1994}.
Furthermore, the fluctuation-induced diamagnetic susceptibility of a superconducting film in the normal state is orders of magnitude smaller than the perfect diamagnetism in the Meissner state, rendering the self-field negligible \cite{Skocpol1975}.

By choosing the local Landau gauge with zero transverse component, a suitable vector potential is $\bm{A}_\mathrm{s} = (0, B u)$, which satisfies $\nabla \times \bm{A}_\mathrm{s} = B \,{\bm{e}_3}$.
From Eq.\ (\ref{eq:LGL_curv_surface_scaled}), the linearized GL equation on the film takes the form:
\begin{equation}\label{eq:gl_cylindrical_segment}
  -\partial_u^2\chi_\mathrm{s} - \left(\partial_v - \mathrm{i} B u\right)^2 \chi_\mathrm{s} - \frac{1}{2R^2} \chi_\mathrm{s} = -\alpha \chi_\mathrm{s},
\end{equation}
where the term $-1/(2R^2)$ corresponds to the dimensionless superconducting GP, derived from $M=1/(2R)$ and $K=0$.
With the origin of the coordinates $(u,v)$ set at the center of the film, the boundary conditions (\ref{eq:BC_curv_surface}) for Eq.\ (\ref{eq:gl_cylindrical_segment}) become:
\begin{equation}\label{eq:bc_cylindrical_segment}
  \partial_u \chi_\mathrm{s}\big|_{u=\pm U/2} = 0,\
  \left(\partial_v - \mathrm{i} B u \right) \chi_\mathrm{s} \big|_{v=\pm V/2} = 0.
\end{equation}

Although Eq.\ (\ref{eq:gl_cylindrical_segment}) resembles the Schr\"{o}dinger equation for a 2D free electron in a magnetic field, the finite boundary conditions (\ref{eq:bc_cylindrical_segment}) preclude a simple analytical solution.
Therefore, we employ the finite element method to numerically determine the critical temperature $\tilde{T}^\ast$, which corresponds to the lowest eigenvalue $-\alpha$.
Before presenting the results, it is useful to estimate the strain in such bent films to justify our focus on geometric effects.
For a mechanically bent film with $d=0.1\xi_0$ and $R \approx 4.17\xi_0$, the maximum strain would be approximately $1.2\%$. Recent studies reveal that such strain levels can induce significant shifts in $T_c$ (e.g., up to $1.5$ K in flexible NbSe$_2$ layers) \cite{henriquez2024modulation,wieteska2019uniaxial}, which could mask the purely geometric GP effect. To unambiguously isolate the curvature-enhanced superconductivity, we propose that the thin film should not be mechanically bent, but rather directly grown or deposited conformally onto a pre-curved cylindrical substrate (e.g., using conformal chemical vapor deposition or 3D FIBID) \cite{cordoba2019three}. In this pristine grown geometry, the film is inherently stress-free, ensuring that the observed $T_c$ enhancement is exclusively driven by the superconducting GP.

\begin{figure}[!h]\centering
  \includegraphics[width=0.9\columnwidth]{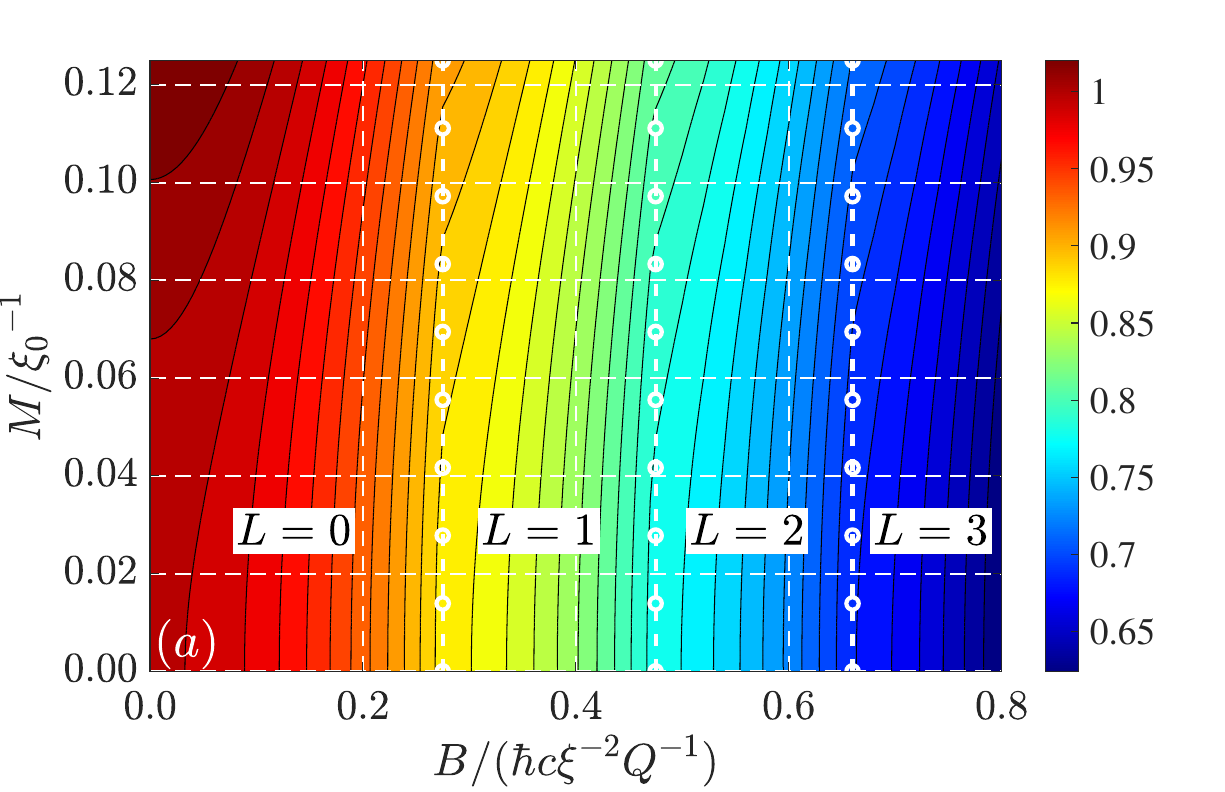}\\%
  \includegraphics[width=0.9\columnwidth]{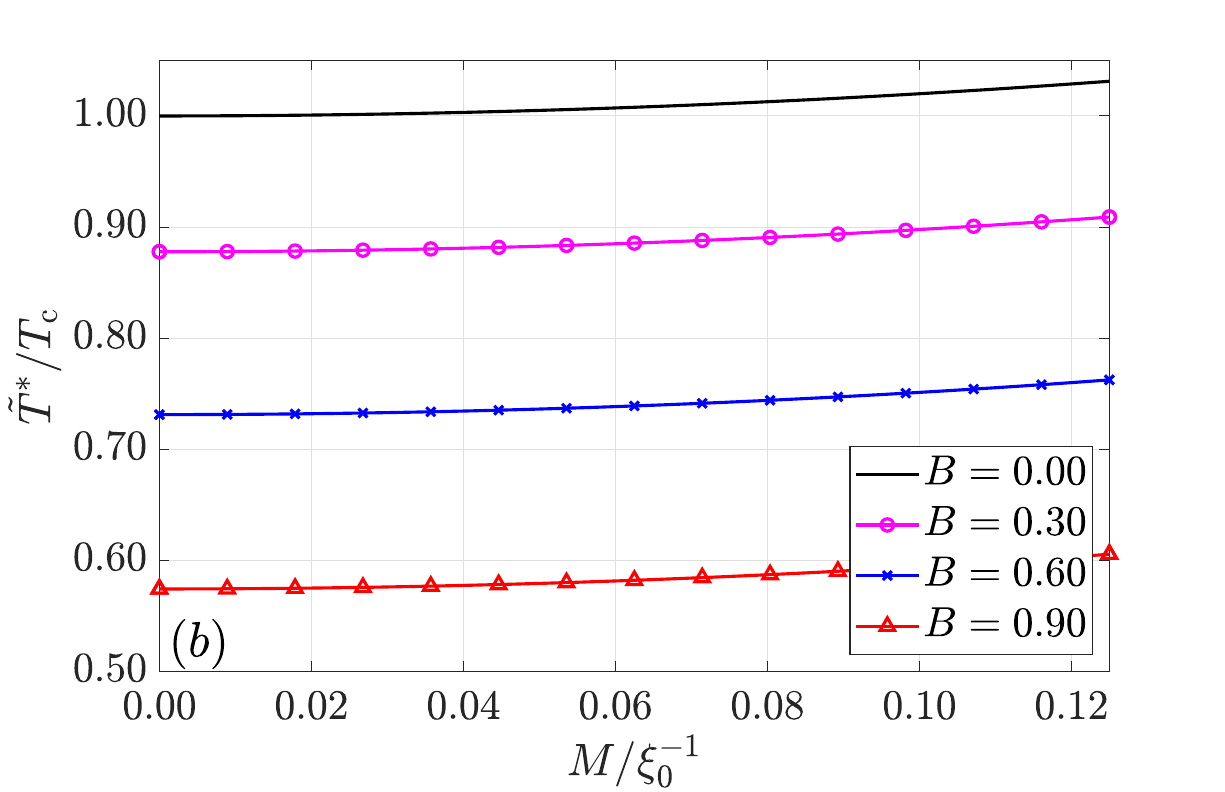}\\%
  \includegraphics[width=0.9\columnwidth]{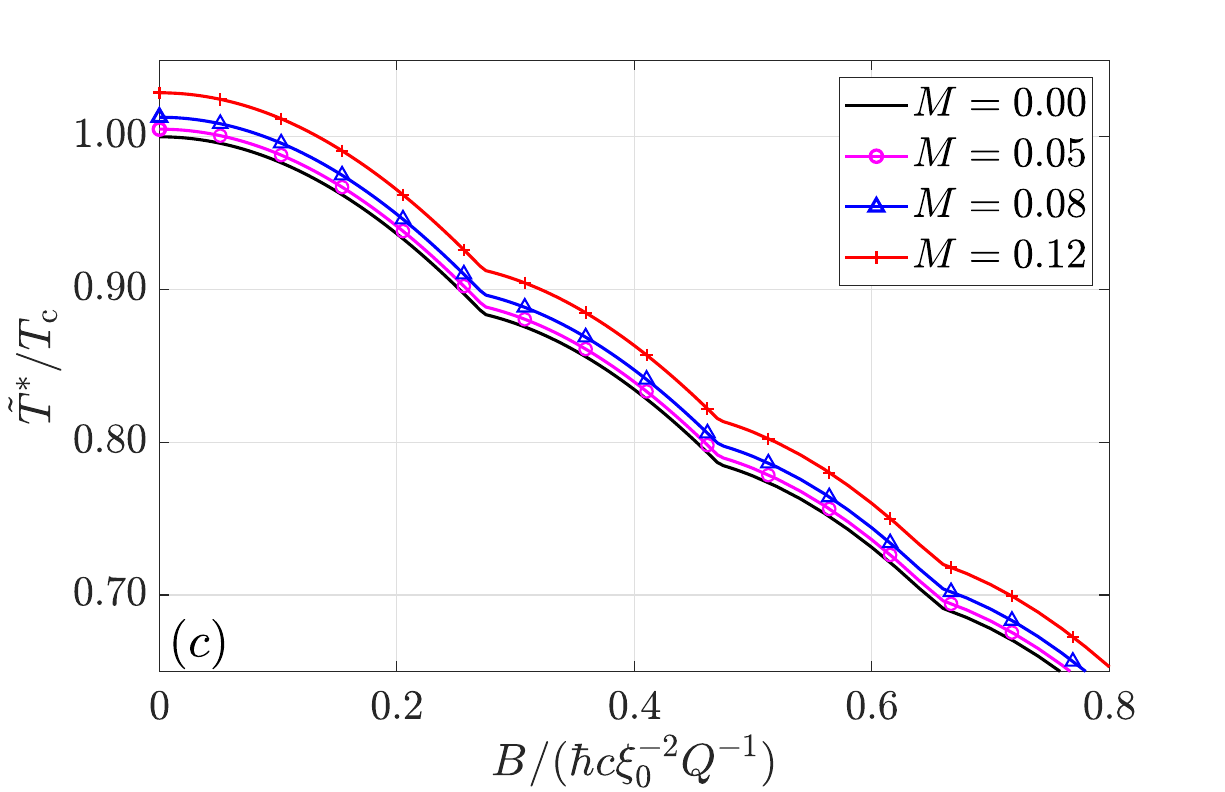}
  \caption{\label{fig:mapOfTc}
    (Color online)
    (a) Phase diagram of the critical temperature $\tilde{T}^\ast$ (in units of $T_\mathrm{c}$) for the cylindrical film model ($U = 8 \xi_0, V = 6 \xi_0$) as a function of mean curvature $M$ ($\xi_0^{-1}$) and magnetic field $B$ ($\hbar c \xi_0^{-2} Q^{-1}$).
    Solid black curves are contour lines of $\tilde{T}^\ast$.
    Dashed lines marked with circles delineate boundaries between regions with different vorticity $L$.
    (b) $\tilde{T}^\ast$ versus curvature $M$ for selected magnetic fields $B=0.0, 0.30, 0.60, 0.90$.
    The quadratic increase confirms the $M^2$ dependence.
    (c) $\tilde{T}^\ast$ versus magnetic field $B$ for curvatures $M=0, 0.05, 0.08, 0.12$.
    This illustrates the competition between curvature-induced enhancement and magnetic field suppression.
  }
\end{figure}

Fig.\ \ref{fig:mapOfTc} (a) displays the phase diagram of the critical temperature $\tilde{T}^\ast$ (in units of $T_\mathrm{c}$) as a function of mean curvature $M$ (in units of $\xi_0^{-1}$) and magnetic field intensity $B$ (in units of $\hbar c \xi_0^{-2} Q^{-1}$), for a film with $U = 8 \xi_0$ and $V = 6 \xi_0$.
Fig.\ \ref{fig:mapOfTc} (b) shows slices of $\tilde{T}^\ast$ at different magnetic field intensities as a function of mean curvature.
Consistent with the analytical prediction in Eq.\ (\ref{eq:Tc_mod}), the geometric effect effectively lowers the kinetic energy of the system, leading to an increase in $\tilde{T}^\ast$.
The numerical results confirm that $\tilde{T}^\ast$ depends quadratically on $M$.
Fig.\ \ref{fig:mapOfTc} (c) presents slices of $\tilde{T}^\ast$ at different curvatures as a function of magnetic field $B$.
$\tilde{T}^\ast$ decreases monotonically with increasing magnetic field and exhibits oscillatory behavior, where each cusp corresponds to a transition between nucleation states characterized by different vorticity $L$ \cite{moshchalkov1995effect}.
By connecting the corresponding cusps in the $\tilde{T}^\ast(B)$ curves for different curvatures (white dashed lines marked with circles in Fig.\ \ref{fig:mapOfTc} (a)), we can map out the phase boundaries between states with different vorticity $L$.
Compared to the flat film ($M=0$), all $\tilde{T}^\ast(B)$ curves for finite $M$ are shifted upward.
Quantitative analysis reveals that these shifts scale as $2M^2$, precisely matching the correction derived from the superconducting GP, $V_\mathrm{g}^\mathrm{s}$.
These findings corroborate that geometric curvature significantly enhances the critical temperature of the film, verifying the theoretical prediction in Eq.\ (\ref{eq:Tc_mod}).
Moreover, we propose that this cylindrical film configuration serves as an ideal platform for the experimental verification of curvature-enhanced superconductivity.
As an additional feature of this open rectangular geometry, the phase boundary $T_c(B)$ shown in Fig.\ \ref{fig:mapOfTc} exhibits a distinct oscillatory behavior. This modulation arises directly from the successive nucleation of vortices entering through the open edges. The system transitions between varying vorticity states, denoted by $L$, as the applied magnetic field provides sufficient energy for the penetrating vortices to overcome a surface edge barrier analogous to the Bean-Livingston barrier \cite{bean1964surface,berdiyorov2005surface}.

\section{Proposed Experimental Validation in Strain-Free Ultracold Atomic Systems \label{sec:Proposed_Experimental}}

Although the numerical simulations in Sec.\ \ref{sec:Curvature-induced_nucleation} predict a curvature-induced enhancement of superconductivity, quantitative verification in solid-state devices is complicated by strain effects that often mask intrinsic geometric contributions \cite{Ortix2011Curvature,Heinrich2025Bending}.
To isolate the predicted superconducting GP within a strain-free environment, we propose utilizing ultracold atomic condensates where the macroscopic wavefunction is governed by the Gross-Pitaevskii equation (GPE), a formal analogue of the GL equation, thereby offering an ideal platform devoid of lattice deformation artifacts.

Realizing this experimental paradigm hinges on satisfying two technical prerequisites: the construction of a curved two-dimensional manifold and the implementation of Neumann-type boundary conditions ($\nabla\psi\cdot\bm{n}=0$).
With regard to the former, shell-shaped Bose-Einstein condensates (BECs) realized in orbital microgravity via radiofrequency dressing exhibit micrometer-scale thicknesses satisfying the thin-layer limit, ($d \ll R$), thereby providing a strain-free and curvature-tunable substrate \cite{carollo2022observation}.
As for the latter, conventional optical ``box traps'', implemented via repulsive blue-detuned lasers \cite{gaunt2013bose} or digital micromirror devices \cite{gauthier2016direct}, impose Dirichlet boundary conditions, yielding a da Costa potential that vanishes identically on a sphere ($V_\mathrm{g} \propto M^2 - K = 0$).
In stark contrast, the imposition of Neumann boundary conditions engenders a non-vanishing superconducting GP that persists on a sphere, scaling as an attractive interaction $V_\mathrm{g}^\mathrm{s} \propto -1/R^2$.
To experimentally engineer this requisite boundary condition, we propose a concentric multi-shell architecture achieved via multi-frequency radiofrequency dressing \cite{harte2018ultracold,lundblad2019shell}, which establishes a radially periodic trapping potential.
Within each unit cell of such a periodic system, the condensate wavefunction in the ground Bloch state inherently satisfies the Neumann boundary condition at both the inner and outer potential barriers due to reflection symmetry.
Consequently, the observation of any curvature-dependent shift in the condensation temperature or chemical potential within a spherical shell would be a smoking-gun signature of the superconducting GP, definitively distinguishing it from da Costa's quantum confinement.
To ensure strictly two-dimensional physical behavior and avoid inter-shell cross-talk, the multi-shell barrier height must be tuned to the deep optical lattice limit to exponentially suppress quantum tunneling. Furthermore, since ultracold atoms are electrically neutral, simulating the magnetic field response governed by Eq.\ (\ref{eq:bc_cylindrical_segment}) necessitates the implementation of artificial gauge fields \cite{dalibard2011colloquium,lin2009synthetic}, which can be engineered through Raman-assisted tunneling or by rotating the condensate trap.

\section{Conclusion \label{sec:Conclusion}}
In this work, we have established a rigorous theoretical framework for the linearized GL equation on curved ultra-thin superconducting films.
By employing the TLQS and adopting a specific gauge transformation to decouple the surface and transverse degrees of freedom, we derived an effective surface equation governing the superconducting order parameter.
A central finding is the identification of the superconducting GP, $V_\mathrm{g}^\mathrm{s} = -\hbar^2(2M^2 - K)/(2\mu)$, which differs from da Costa's GP due to the transverse Neumann boundary condition inherent to the superconducting order parameter.
From the perspective of the GL free energy, we elucidated the mechanism of curvature-induced superconductivity enhancement.
The negative GP acts as an effective attractive interaction that reduces the transverse ground-state kinetic energy required to confine Cooper pairs within the curved film.
Consequently, the superconducting state can nucleate at temperatures higher than the critical temperature of a corresponding planar film, with a relative shift proportional to $\xi_0^2(2M^2 - K)$.
Importantly, this geometric enhancement arises from macroscopic quantum confinement and is distinct from microscopic strain effects that modify the electron-phonon coupling.
We numerically validated these predictions by simulating the phase transition of a rectangular film bent onto a cylindrical surface under a perpendicular magnetic field.
The calculated phase boundary exhibits oscillatory behavior corresponding to transitions between nucleation states with different vorticity, rather than simple vortex entry dynamics.
Most significantly, the critical temperature was found to increase quadratically with mean curvature, in precise quantitative agreement with our theoretical prediction.
Beyond the domain of curved superconducting films, we outline a strain-free validation scheme using ultracold atomic condensates.
This proposed configuration utilizes nested superfluid shells to enforce Neumann boundary conditions, which would isolate the non-vanishing superconducting GP on a sphere.
Distinct from the vanishing da Costa potential, this effect would offer an unambiguous signature of curvature-induced enhancement.
We hope that these results will stimulate experimental verification of the superconducting GP, utilizing either the curved thin-film geometries modeled here or the proposed ultracold atomic condensate systems.

\section{Acknowledgments}
This work is supported by the Science and Technology Program of Guangxi, China (Grant No.\ AD19110126), the National Natural Science Foundation of China (Grants No.\ 12147103, No.\ 12004149 and No.\ 12475019), Natural Science Foundation of Shandong Province of China (Grants No.\ ZR2020MA091 and ZR2020QA062) and National Lab of Solid State Micro-structure of Nanjing University (Grants No.\ M35040 and No.\ M35053).

\appendix
\section{Exact Solution of the Transverse Linearized GL Equation \label{app:Exact_Solution}}

In this appendix, we solve the transverse eigenvalue problem governed by Eq.\ (\ref{eq:LGL_curv_transverse}) under the boundary condition (\ref{eq:bc4_normal}).
Defining $\lambda \equiv 2\mu \alpha_\mathrm{t} / \hbar^2$, the general solution to the differential equation takes the form $\chi_\mathrm{t}(q^3) = c_1 \mathrm{e}^{\sqrt{\lambda} q^3} + c_2 \mathrm{e}^{-\sqrt{\lambda} q^3}$.
Imposing the Neumann-type boundary condition $[(\partial_3 - M)\chi_\mathrm{t}]|_{q^3=\pm d/2} = 0$ results in the following system of homogeneous linear equations for the coefficients $c_1$ and $c_2$:
\begin{align}
  & (\sqrt{\lambda} - M) \,\mathrm{e}^{\sqrt{\lambda} d/2} c_1 - (\sqrt{\lambda} + M) \,\mathrm{e}^{-\sqrt{\lambda} d/2} c_2 = 0, \nonumber\\
  & (\sqrt{\lambda} - M) \,\mathrm{e}^{-\sqrt{\lambda} d/2} c_1 - (\sqrt{\lambda} + M) \,\mathrm{e}^{\sqrt{\lambda} d/2} c_2 = 0.
\end{align}
For non-trivial solutions ($c_1, c_2$ not both zero) to exist, the determinant of the coefficient matrix must vanish, leading to the characteristic equation:
\begin{equation}
  (\lambda - M^2) \sinh(\sqrt{\lambda} d) = 0.
\end{equation}
This condition allows for two distinct classes of eigenvalues.
The first class arises from the roots of $\sinh(\sqrt{\lambda} d) = 0$, implying $\lambda_n = -(n\pi/d)^2$ for non-zero integers $n$.
These modes correspond to high-energy transverse standing waves with kinetic energy scaling as $d^{-2}$, which diverges in the thin-film limit ($d \to 0$).
The second solution corresponds to $\lambda = M^2$, which yields a finite eigenvalue $-\alpha_\mathrm{t} = -\hbar^2 M^2 / (2\mu)$.
The associated eigenstate is given by $\chi_\mathrm{t}(q^3) \propto \exp(M q^3)$.
Since this geometry-induced mode represents the lowest energy state (energetically favorable) for superconducting nucleation in the regime $d \ll R$, we retain this solution for the derivation of the effective surface equation.

\bibliographystyle{apsrev4-2}
\bibliography{paperPraSub}

\end{document}